\documentclass{article}
\usepackage{spconf,amsmath,graphicx}
\usepackage{setspace}
\usepackage{amsfonts}
\pdfoutput=1
\usepackage{algorithmic}
\usepackage{array}
\usepackage[caption=false,font=normalsize,labelfont=sf,textfont=sf]{subfig}
\usepackage{textcomp}
\usepackage{stfloats}
\usepackage{url}
\usepackage{verbatim}
\usepackage{graphicx}
\usepackage{cite}
\usepackage{booktabs}
\usepackage{multirow}
\usepackage{threeparttable}
\usepackage{footnote}
\pdfoutput=1
\title{Efficient Polyp Segmentation Via Integrity Learning}
\name{Ziqiang Chen, Kang Wang\sthanks{Ziqiang Chen and Kang Wang contributed equally to this work.}, Yun Liu\sthanks{Corresponding author.}}
\address{School of Basic Medical Sciences, Fudan University, China}
\begin{document}
\ninept
\maketitle

\begin{abstract}
Accurate polyp delineation in colonoscopy is crucial for assisting in diagnosis, guiding interventions, and treatments. However, current deep-learning approaches fall short due to integrity deficiency, which often manifests as missing lesion parts. This paper introduces the integrity concept in polyp segmentation at both macro and micro levels, aiming to alleviate integrity deficiency. Specifically, the model should distinguish entire polyps at the macro level and identify all components within polyps at the micro level. Our \textbf{I}ntegrity \textbf{C}apturing \textbf{Polyp} \textbf{Seg}mentation (IC-PolypSeg) network utilizes lightweight backbones and 3 key components for integrity ameliorating: 1) Pixel-wise feature redistribution (PFR) module captures global spatial correlations across channels in the final semantic-rich encoder features. 2) Cross-stage pixel-wise feature redistribution (CPFR) module dynamically fuses high-level semantics and low-level spatial features to capture contextual information. 3) Coarse-to-fine calibration module combines PFR and CPFR modules to achieve precise boundary detection. Extensive experiments on 5 public datasets demonstrate that the proposed IC-PolypSeg outperforms 8 state-of-the-art methods in terms of higher precision and significantly improved computational efficiency with lower computational consumption. IC-PolypSeg-EF0 employs 300 times fewer parameters than PraNet while achieving a real-time processing speed of 235 FPS. Importantly, IC-PolypSeg reduces the false negative ratio on five datasets, meeting clinical requirements.
\end{abstract}

\section{Introduction}
\label{sec:intro}

% These guidelines include complete descriptions of the fonts, spacing, and
% related information for producing your proceedings manuscripts. Please follow
% them and if you have any questions, direct them to Conference Management
% Services, Inc.: Phone +1-979-846-6800 or email
% to \\\texttt{papers@2024.ieeeicassp.org}.

Colorectal cancer (CRC) is one of the leading cancers, causing over 1.93 million new cases and 0.94 million estimated deaths globally in 2020~\cite{sung_global_2021}. Colon polyps, particularly adenomas with high-grade dysplasia, are the most common cause of CRC. Regular early diagnosis and intervention are essential in colon cancer prevention and treatment. Colonoscopy is regarded as the gold standard for screening colon adenomas and colorectal cancer. However, due to limited healthcare resources, there is time pressure on endoscopy screening, leading to a high rate of missed polyps during the entire procedure, reported to be between 20\% and 47\%~\cite{leufkens2012factors}. Hence, computer-aided tools are indispensable in supporting endoscopists and enhancing polyp detection during endoscopy screening.    

Automated polyp segmentation of endoscopy images is challenging, as colorectal polyps vary extensively in size and shape. To remedy this, many approaches have been proposed. 
Recently, Deep learning (DL) has emerged as a powerful tool. Learning-based methods, such as Fully Convolutional Networks (FCN)~\cite{long2015fully} and UNet~\cite{unet}, are widely employed in image segmentation and have further evolved to cater specifically to polyp segmentation. 

\begin{figure}
  \centering
  \includegraphics[width=1\linewidth]{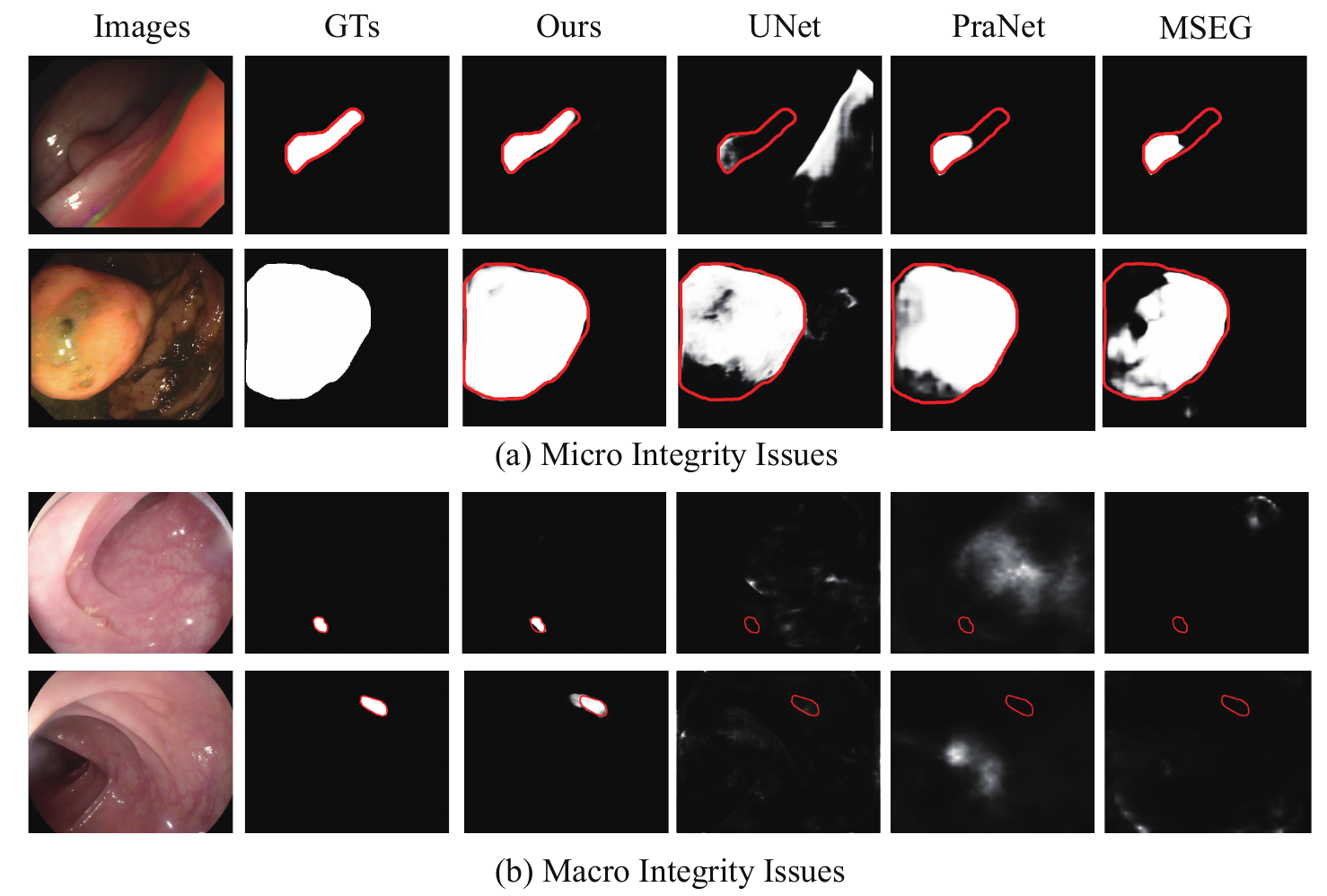}
  \caption{Visual examples of integrity issues (\textit{i.e.}, (a) micro integrity and (b) macro integrity) from polyp segmentation. These columns show the input images, ground truth images, and results of our IC-PolypSeg-EF0, UNet, PraNet, and MSEG, separately.}
  \label{integrity_issue}
  \vspace{-0.6 cm}
\end{figure}

Current advances in learning-based methods largely benefit from leveraging multi-scale feature aggregation, contextual learning, and boundary-aware mechanisms. Multi-scale feature aggregation mechanisms are widely used in polyp segmentation and enhance the features from different layers to distinguish polyp areas of various sizes. For example, Zhao \textit{et al.}~\cite{apsms} proposed a multi-scale subtraction network (MSNet) to segment polyp from colonoscopy images with several subtraction units. Similarly, Song \textit{et al.} used multi-scale learning modules with attention settings to generate better polyp feature representation \cite{song2022attention}.

Contextual learning is another key component in polyp segmentation. Zhong \textit{et al.} designed PolypSeg \cite{polypseg} with an adaptive scale context module to aggregate multi-scale contextual features. Zhang \textit{et al.} proposed ACSNet \cite{acsnet} with an Adaptive Selection Module (ASM) for contextual aggregation and selection.

To accurately locate polyp boundaries, boundary-aware learning methods have been proposed to direct networks' attention toward the contours that separate polyps from the background. Fang \textit{et al.}~\cite{fang2019selective} proposed selective feature aggregation with boundary constraint loss to produce the final result. Cheng \textit{et al.}~\cite{cheng2021learnable} designed the Learnable Oriented Derivative Network (LOD) to extract boundary features.

While the previously mentioned methods enhance polyp segmentation performance in various aspects, they do not address the issue of integrity in polyp segmentation. As illustrated in Fig.~\ref{integrity_issue}, we refer \cite{zhuge2022salient} and define the integrity issue in polyp segmentation at two levels. At the macro level, the methods need accurately distinguish all polyp regions within the provided images. At the micro level, the models should correctly identify all the constituent parts within specific polyp areas.

To enhance the ability of integrity learning and precision of polyp segmentation, we propose the IC-PolypSeg for accurate and efficient polyp segmentation. IC-PolypSeg comprises three main components. Firstly, a pixel-wise feature redistribution (PFR) module in the final encoder stage captures pixel relationships across channels to better preserve integrity. Secondly, various cross-stage pixel-wise feature redistribution (CPFR) modules capture contextual information at the pixel level to capture contexual information. Finally, a coarse-to-fine calibration module combines PFR and CPFR modules to produce precise boundaries.

\section{Methods}
\label{sec:method}

\begin{figure}
\centering
  \includegraphics[width=1\linewidth]{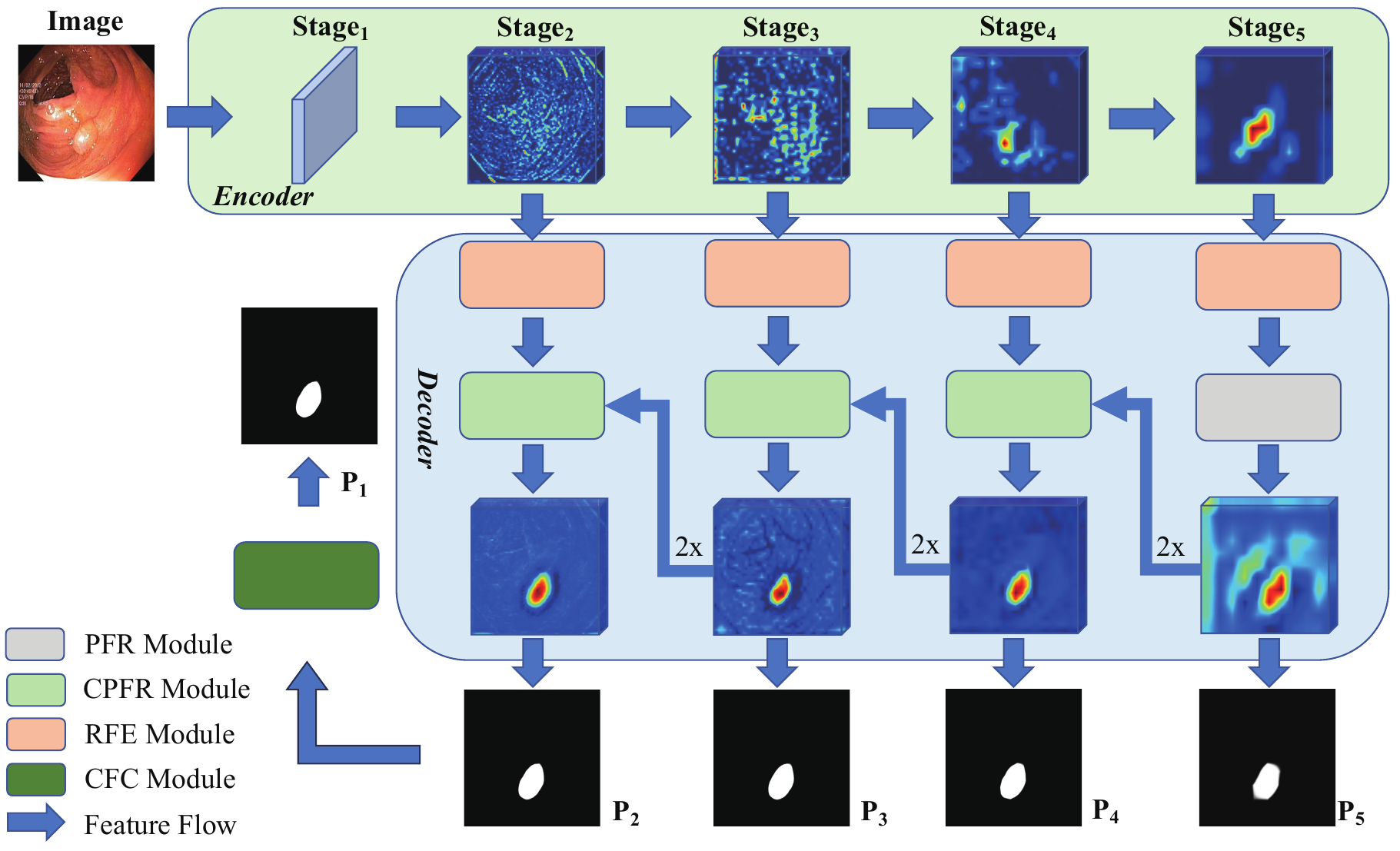}
  \caption{Overview architecture of IC-PolypSeg.}
\label{overview}
\vspace{-0.6 cm}
\end{figure}

\subsection{Framework}

As illustrated in Fig.~\ref{overview}, our proposed method employs a UNet-like architecture. The encoder serves for multi-level feature extraction, and the decoder integrates these features, incorporating deep supervision and adaptive pixel loss~\cite{tracer} to generate segmentation maps. To enhance the network's flexibility and applicability, the universal encoder can accommodate various feature extractors with different backbones. We denote the encoder features as ${F}_{enc}=\{\mathbf{F}_{enc}^{(1)}, \mathbf{F}_{enc}^{(2)}, \mathbf{F}_{enc}^{(3)}, \mathbf{F}_{enc}^{(4)}, \mathbf{F}_{enc}^{(5)}\}$, excluding ${F}_{enc}^{(1)}$ given its large spatial size.   

Subsequently, we introduce receptive field expanding (RFE) modules to enlarge the receptive fields of encoder features. Additionally, pixel-wise feature redistribution (PFR) and cross-stage pixel-wise feature redistribution (CPFR) modules are incorporated to enhance pixel-wise representations across decoder stages. As depicted in Fig.~\ref{overview}, we generate prediction $P_{i}$ from $\text{stage}_{i}$ (for $i\in\{2,3,4,5\}$). Finally, the coarse-to-fine calibration (CFC) module refines the coarse prediction $P_{2}$ from the last decoder stage.

\subsection{Receptive Field Expanding Module}
The RFE module follows a universal design inspired by Inception-like modules, employing four branches with different convolutional kernel sizes to expand the receptive fields of encoder features. The procedure of an RFE module can be summarized as follows:
\vspace{-2mm}
% \mathcal{X}_{a s y}(\mathbf{I})
\begin{small}
\begin{equation}
\begin{aligned}
F_{rfe}^{i} = \mathcal{X}_{rfe} * Concat(\mathcal{X}_{1}(F_{enc}^{i}), \mathcal{X}_{3}(F_{enc}^{i}), \mathcal{X}_{5}(F_{enc}^{i}), \mathcal{X}_{7}(F_{enc}^{i})),
\end{aligned}
\end{equation}
\end{small}
Where * denotes the 2D convolutional operator, and $F_{ref}^{i}$ is the output features from the RFE module. Let $\mathcal{X}_{i}$, $i \in {1, 2, 3, 4}$ denote the different weights matrices in the RFE module, and $Concat(\cdot)$ denotes the concatenation operation. Specifically, $\mathcal{X}_{1}$ is a 1 × 1 convolutional kernel $K_{1 \times 1}$. $\mathcal{X}_{j}$, $j \in {3, 5, 7}$, all contains three layers: one with a 1 × 1 kernel $K_{1 \times 1}$, one with a horizontal 1 × j kernel $K_{1 \times j}$, and one with a vertical j × 1 kernel $K_{j \times 1}$. Finally, $\mathcal{X}_{rfe}$ is a 1 × 1 convolutional kernel that reduces the number of channels to $C$.

\subsection{Pixel-wise Feature Redistribution Module}

\begin{figure}
\centering
\includegraphics[width=1\linewidth]{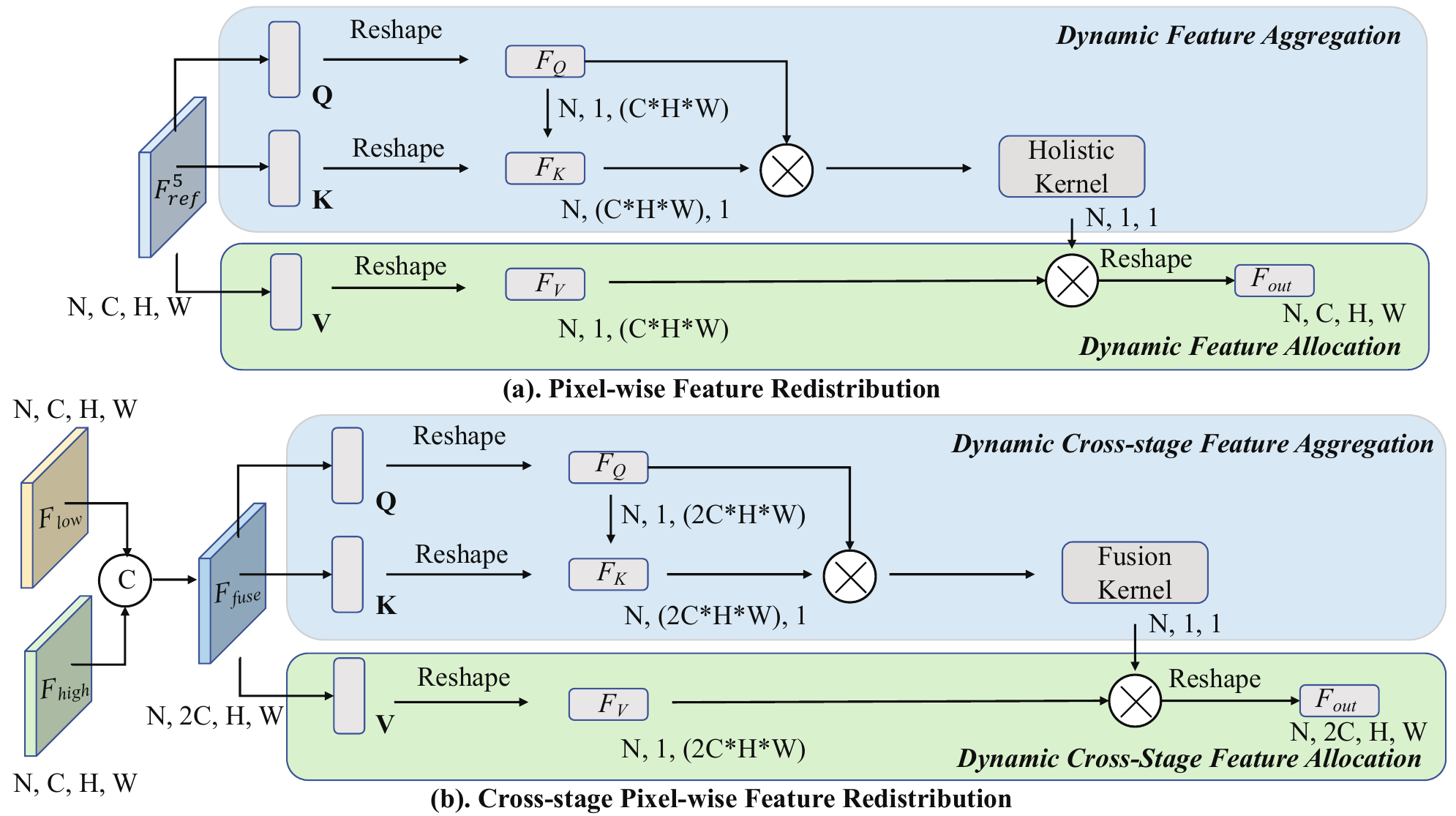}
\caption{Details of the proposed PFR and CPFR modules. }
\label{vis}
\end{figure}

As depicted in Figure~\ref{vis}(a), the PFR module comprises two phases: dynamic feature aggregation and dynamic feature allocation.

In the dynamic feature aggregation phase, PFR module fuses features via \textbf{Q} and \textbf{K} features. Given a local feature $F_{ref}^{5} \in \mathbb{R}^{N \times C \times H \times W}$ from the last encoder, we first feed it into two $1\times1$ convolutional layers to generate two feature maps \textbf{Q} and  \textbf{V}, respectively, where $\{\textbf{Q}, \textbf{K}\} \in \mathbb{R}^{N \times C \times H \times W}$. Then, we reshape the \textbf{Q} to $F_{Q}$ and \textbf{K} to $F_{K}$, where $F_{Q} \in \mathbb{R}^{N \times 1 \times (C \times H \times W)}$ and $F_{K} \in \mathbb{R}^{N  \times (C \times H \times W) \times 1}$. After that we implement a batch matrix multiplication between the $F_{Q}$ and $F_{K}$ to generate holistic correlation metric $F_{H}$, where $F_{H} \in \mathbb{R}^{N \times 1 \times 1}$. 

In the dynamic feature allocation phase, we feed $F_{ref}^{5} \in \mathbb{R}^{N \times C \times H \times W}$ into a $1\times1$ convolutional layer to generate feature \textbf{V} and reshape it to $F_{V} \in \mathbb{R}^{N \times 1 \times (C \times H \times W)}$. Then, we perform a batch metric multiplication between holistic metric $F_{H}$ and $F_{V}$ and reshape the result to $F_{pfr} \in \mathbb{R}^{N \times C \times H \times W}$. 

\subsection{Cross-Stage Pixel-wise Feature Redistribution Module}
After obtaining $F_{pfr} \in \mathbb{R}^{N \times C \times H \times W}$, we upsample it by a factor of two and deliver it to the adjacent layer with CPFR module. The CPFR module takes two input features: the output of the corresponding RFE module, and $F_{pfr}$ or the result from the previous CPFR module. The CPFR module aggregates high-level semantic features and low-level spatial information to generate a cross-stage holistic feature kernel ($F_{Hol}^{C} \in \mathbb{R}^{N \times 1 \times 1}$) through $F_{Q} \in \mathbb{R}^{N \times 1 \times (2C \times H \times W)}$ and $F_{K} \in \mathbb{R}^{N \times (2C \times H \times W) \times 1}$ in the dynamic cross-stage feature aggregation phase. Meanwhile, the CPFR module generates a feature $F_{V}$ and performs batch matrix multiplication between $F_{Hol}^{C}$ and $F_{V}$, reshaping the result to $F_{G} \in \mathbb{R}^{N \times 2C \times H \times W}$ during dynamic cross-stage feature allocation. Finally, we apply a $1 \times 1$ convolutional layer to reduce the channel dimension and generate the output $F_{cpfr} \in \mathbb{R}^{N \times C \times H \times W}$.

\subsection{Coarse-to-fine Calibration Module}

Coarse-to-fine calibration module is designed as a plugin residual block to refine the coarse predicted polyp segmentation maps (as shown $P_{2}$ in Fig.~\ref{overview}) by learning the residuals $P_{res}$ between the coarse prediction maps and the ground truth as:  $P_{1} = P_{2} + P_{res}$

\begin{figure}
\centering
  \includegraphics[width=0.8\linewidth]{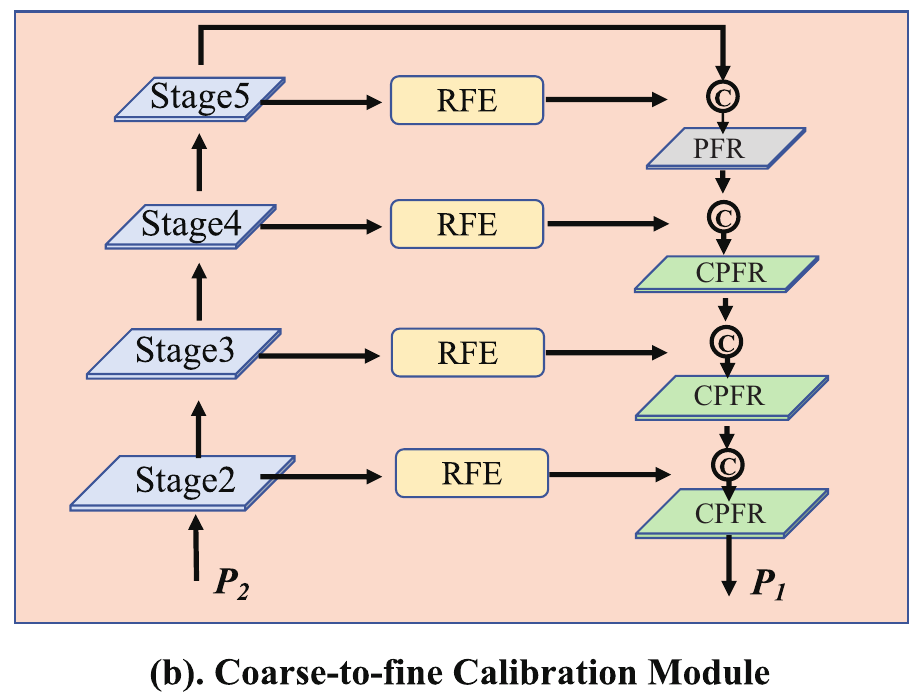}
\caption{Details of the Coarse-to-fine calibration module.}
\label{cfcm}
\end{figure}

As depicted in Figure~\ref{cfcm}, the CFC module is designed following the classical Encoder-Decoder architecture \cite{qin2019basnet}, integrating RFE, PFR, and CPFR modules into the decoder phase. We have configured four stages for both the encoder and decoder phases. Each encoder stage comprises a $3 \times 3$ convolutional layer with 32 filters, along with batch normalization and ReLU activation. To achieve downsampling, we use non-overlapping 2 × 2 max pooling. RFE modules are first used to expand the receptive field of the encoder features. The encoder features then pass through the PFR and CPFR modules sequentially. During inference, the output of the CFC module serves as the final prediction map.

\section{Experiments}
\subsection{Datasets and Metrics}
Five public datasets are encompassed for algorithm accuracy and effectiveness evaluation, including CVC-ClinicDB~\cite{cvc-clinicdb}, Kvasir-SEG~\cite{Kvasir-SEG}, ColonDB~\cite{ColonDB}, Endoscene~\cite{Endoscene} and ETIS~\cite{ETIS}. Following the experimental settings in PraNet~\cite{pranet}, we utilize training datasets from the Kvasir-SEG and ClinicDB datasets, while the test sets are composed of ColonDB, ETIS, and EndoScene. Our evaluation of IC-PolypSeg and other state-of-the-art (SOTA) methods are based on four metrics: Dice coefficient (Dice), intersection over union (IoU), mean absolute error (MAE), and false negative ratio (FNR).

\subsection{Implementation Details}
All experiments and model construction are implemented on a Linux platform equipped with four NVIDIA RTX2080Ti GPUs and the PyTorch 1.10 framework. All the inputs are resized to the scale of $352 \times 352$ and employ a multi-scale ratio from $\{0.75, 1, 1.25\}$. The encoder parameters are initialized from EfficientNet-B0$ \sim $B7 series networks. We use the AdamW \cite{adamw} optimizer to update the network parameters. The learning rate and weight decay are both set to 1e-4. Further, The batch size is set to 16 and the maximum number of training epochs is set to 100. Besides,an early stopping operation was applied when the validation loss did not decrease after 10 epochs. For testing, we only resize the input images to $352 \times 352$ without any post-processing.

\subsection{Quantitative Comparison with the SOTAs}
We compare IC-PolypSeg with 8 recent SOTA methods, including UNet~\cite{unet}, UNet++~\cite{unet++}, SFA~\cite{sfa}, MSEG~\cite{hardnet}, DCRNet~\cite{dcrn}, ACSNet~\cite{acsnet}, PraNet~\cite{pranet}, SANet~\cite{sanet}.

\subsubsection{Analysis of Learning Ability}
We employ the ClinicDB and Kvasir-SEG datasets for training, and accordingly, utilizes the corresponding test datasets to evaluate their performance. As shown in Tab.~\ref{seen_dataset}, our methods with EfficientNet backbones achieve competitive results with other SOTAs.

\vspace{-5mm}
\begin{table}[htbp]
  \centering
  \caption{Quantitative results of the seen datasets, Kvasir-SEG and ClinicDB. The best results are in boldface. The EF0 $\sim$ EF7 denotes EfficientNet-B0 $\sim$ EfficientNet-B7. $\uparrow$ and $\downarrow$ denote larger and smaller is better, respectively.}
  \resizebox{\linewidth}{!}{
    \begin{tabular}{c|ccc|ccc}
    \toprule
    \multirow{2}[4]{*}{Model} & \multicolumn{3}{c|}{Kvasir-SEG} & \multicolumn{3}{c}{ClinicDB} \\
\cmidrule{2-7}          & mDice $\uparrow$ & mIoU $\uparrow$  & MAE $\downarrow$   & mDice $\uparrow$ & mIoU $\uparrow$  & MAE $\downarrow$ \\
    \midrule
    U-Net & 0.818 & 0.746 & 0.055 & 0.823 & 0.755 & 0.019 \\
    Unet++ & 0.821 & 0.743 & 0.048 & 0.794 & 0.729 & 0.022 \\
   	SFA   & 0.723 & 0.611 & 0.075 & 0.700   & 0.607 & 0.042 \\
    MSEG  & 0.897 & 0.839 & 0.028 & 0.909 & 0.864 & 0.007 \\
    DCRNet  & 0.886 & 0.825 & 0.035 & 0.896 & 0.844 & 0.01 \\
    ACSNet & 0.898 & 0.838 & 0.032 & 0.882 & 0.826 & 0.011 \\
    PraNet & 0.898 & 0.840  & 0.030  & 0.899 & 0.849 & 0.009 \\
    SANet & 0.904 & 0.847 & 0.028 & 0.902 & 0.859 & 0.012 \\
    \midrule
    IC-PolypSeg-EF0 & 0.900   & 0.841 & 0.029 & 0.921 & 0.870  & 0.009 \\
    IC-PolypSeg-EF1 & 0.904 & 0.846 & 0.027 & 0.934 & 0.885 & 0.007 \\
    IC-PolypSeg-EF2 & 0.907 & 0.849 & 0.029 & 0.924 & 0.875 & 0.007 \\
    IC-PolypSeg-EF3 & 0.907 & 0.852 & 0.029 & 0.932 & 0.881 & 0.008 \\
    IC-PolypSeg-EF4 & 0.898 & 0.840  & 0.031 & 0.935 & 0.886 & 0.007 \\
    IC-PolypSeg-EF5 & 0.897 & 0.841 & 0.029 & 0.934 & 0.883 & 0.007 \\
    IC-PolypSeg-EF6 & 0.906 & 0.851 & 0.028 & 0.937 & \textbf{0.890} & 0.007 \\
    IC-PolypSeg-EF7 & \textbf{0.910} & \textbf{0.859} & \textbf{0.026} & \textbf{0.938} & \textbf{0.890} & \textbf{0.007} \\
    \bottomrule
    \end{tabular}%
    }
  \label{seen_dataset}%
\end{table}%

\vspace{-5mm}

\subsubsection{Analysis of Generalization Ability.}
To verify the generalization performance of our proposed methods, we evaluate them on three unseen datasets, namely ETIS, ColonDB and EndoScene. The Tab.~\ref{unseen_dataset} shows the results of our methods in these three datasets.

\vspace{-5mm}

\begin{table}[htbp]
  \centering
  \caption{Quantitative results of the unseen datasets, ColonDB, ETIS and EndoScene datasets.}
  \resizebox{\linewidth}{!}{
    \begin{tabular}{c|ccc|ccc|ccc}
    \toprule
    \multirow{2}[4]{*}{Model} & \multicolumn{3}{c|}{Colon-DB} & \multicolumn{3}{c|}{ETIS} & \multicolumn{3}{c}{EndoScene} \\
\cmidrule{2-10}          & mDice $\uparrow$  & mIoU $\uparrow$   & MAE  $\downarrow$   & mDice $\uparrow$  & mIoU $\uparrow$   & MAE  $\downarrow$ & mDice $\uparrow$  & mIoU $\uparrow$   & MAE  $\downarrow$ \\
    \midrule
   U-Net & 0.512 & 0.444 & 0.061 & 0.398 & 0.335 & 0.036 & 0.710  & 0.627 & 0.847 \\
   Unet++ & 0.483 & 0.410  & 0.064 & 0.401 & 0.344 & 0.035 & 0.707 & 0.624 & 0.834 \\
   SFA   & 0.469 & 0.347 & 0.094 & 0.297 & 0.217 & 0.109 & 0.467 & 0.329 & 0.644 \\
   MSEG  & 0.735 & 0.666 & 0.038 & 0.700   & 0.630  & 0.015 & 0.874 & 0.804 & 0.948 \\
   DCRNet & 0.704 & 0.631 & 0.052 & 0.556 & 0.496 & 0.096 & 0.856 & 0.788 & 0.943 \\
   ACSNet & 0.716 & 0.649 & 0.039 & 0.578 & 0.509 & 0.059 & 0.863 & 0.787 & 0.939 \\
   PraNet & 0.712 & 0.640  & 0.043 & 0.628 & 0.567 & 0.031 & 0.871 & 0.797 & 0.950 \\
   SANet & 0.753 & 0.670  & 0.043 & 0.750  & 0.654 & 0.015 & 0.888 & 0.815 & 0.962 \\
    \midrule
    IC-PolypSeg-EF0 & 0.753 & 0.674 & 0.035 & 0.707 & 0.617 & 0.021 & 0.892 & 0.822 & 0.965 \\
    IC-PolypSeg-EF1 & 0.768 & 0.683 & 0.034 & 0.720  & 0.634 & 0.016 & 0.900   & 0.832 & 0.972 \\
    IC-PolypSeg-EF2 & 0.796 & 0.713 & \textbf{0.028} & 0.736 & 0.648 & 0.015 & \textbf{0.909} & \textbf{0.844} & \textbf{0.977} \\
    IC-PolypSeg-EF3 & 0.776 & 0.702 & 0.032 & 0.719 & 0.637 & 0.024 & 0.905 & 0.840  & 0.972 \\
    IC-PolypSeg-EF4 & 0.792 & 0.718 & 0.032 & \textbf{0.774} & \textbf{0.692} & 0.016 & 0.904 & 0.839 & 0.975 \\
    IC-PolypSeg-EF5 & 0.789 & 0.710  & 0.031 & 0.761 & 0.676 & 0.014 & 0.904 & 0.837 & 0.974 \\
    IC-PolypSeg-EF6 & \textbf{0.807} & \textbf{0.729} & 0.029 & 0.756 & 0.672 & 0.023 & 0.902 & 0.836 & 0.973 \\
    IC-PolypSeg-EF7 & 0.799 & 0.728 & 0.030  & 0.758 & 0.670  & 0.015 & 0.907 & 0.846 & 0.969 \\
    \bottomrule
    \end{tabular}%
    }
  \label{unseen_dataset}%
\end{table}%

\vspace{-3mm}
\begin{table}[htbp]
  \centering
  \caption{Computational complexity comparisons with others.}
  \resizebox{\linewidth}{!}{
    \begin{tabular}{c|cccc}
    \toprule
    Methods & Backbone & Params(M) $\downarrow$ & MACs(G) $\downarrow$ & \multicolumn{1}{c}{FPS} $\uparrow$ \\
    \midrule
    PraNet & ResNet34 & 32.55 & 13.11 & 50 \\
    ACSNet & Res2Net & 29.45 & 21.61 & 42 \\
    Hardnet-MSEG & Res2Net & 18.45 & 11.39 & 88 \\
    SANet & Res2Net & 23.89 & 11.28 & 49 \\
    \midrule
    IC-PolypSeg-EF0 & EfficientNet-b0 & \textbf{0.67} & \textbf{2.94} & $\sim$ \textbf{235} \\
    IC-PolypSeg-EF1 & EfficientNet-b1 & 0.69 & 2.96  & $\sim$181 \\
    IC-PolypSeg-EF2 & EfficientNet-b2 & 0.70   & 2.96  & $\sim$168 \\
    IC-PolypSeg-EF3 & EfficientNet-b3 & 0.73 & 2.99  & $\sim$146 \\
    IC-PolypSeg-EF4 & EfficientNet-b4 & 0.78 & 3.02  & $\sim$121 \\
    IC-PolypSeg-EF5 & EfficientNet-b5 & 0.84 & 3.06  & $\sim$105 \\
    IC-PolypSeg-EF6 & EfficientNet-b6 & 0.91  & 3.10   & $\sim$95 \\
    IC-PolypSeg-EF7 & EfficientNet-b7 & 1.01  & 3.20   & $\sim$78 
    \\
    \midrule

    \end{tabular}%
    }
  \label{efficiency}%
  \vspace{-6mm}
\end{table}%

\subsubsection{Analysis of Computational Complexity}
In our study, we evaluate the efficiency of our proposed frameworks using model parameters, the number of multiply-accumulate operations (MACs), and GPU inference time (FPS). To ensure fair comparisons, the speed is tested on a NVIDIA 
 RTX2080Ti GPU with a full image resolution of $352 \times 352$. As shown in Table~\ref{efficiency}, our methods achieve competitive performance with less parameters, leveraging the EfficientNet series networks.

\vspace{-4mm}
\subsection{Integrity Analysis}
\begin{figure}[htbp]
\centering
\includegraphics[width=0.9\linewidth]{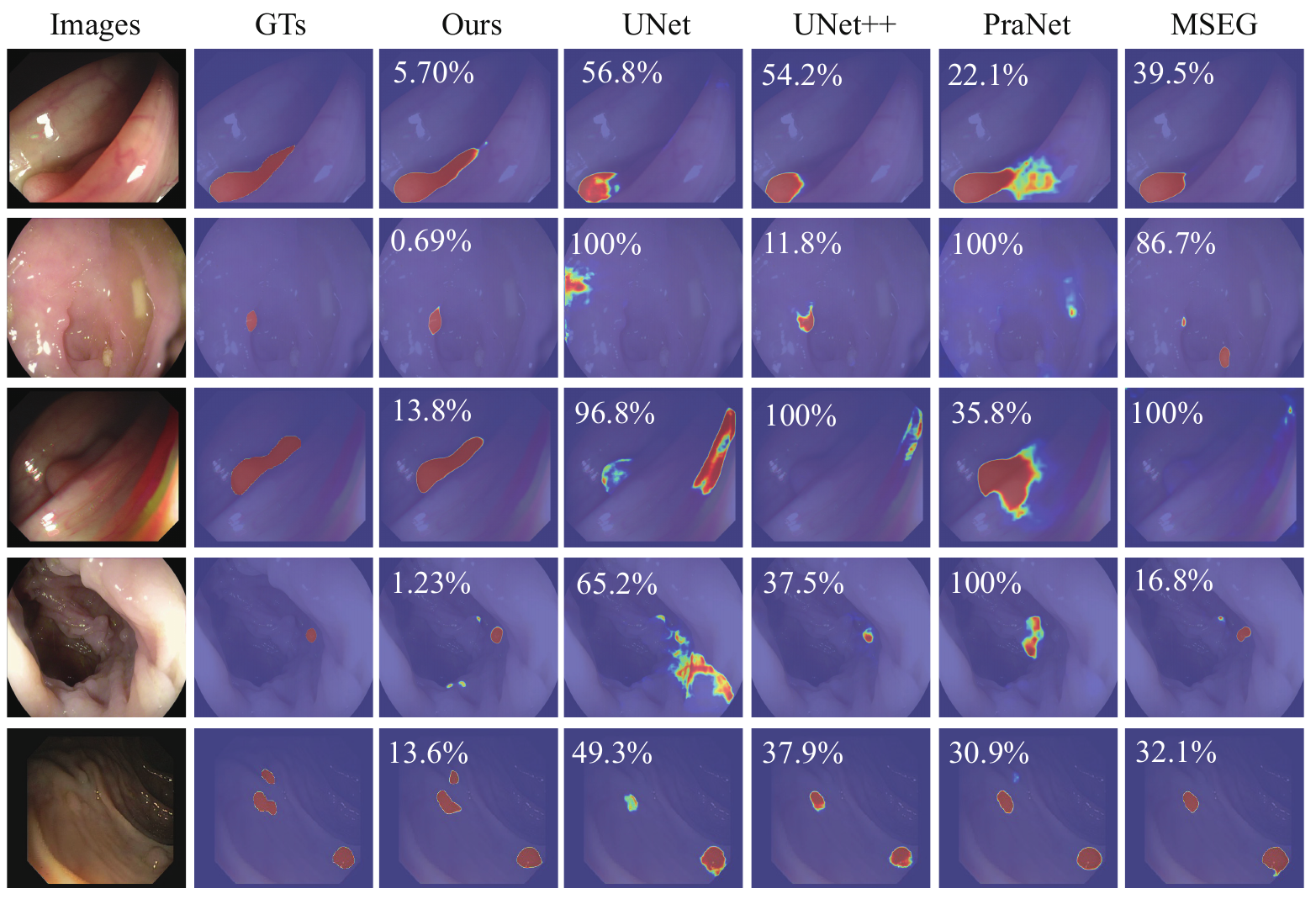}
\caption{Integrity comparisons at different false negative ratios (FNR) of different methods. Best viewed in color.}
\label{intergrity_img}
\vspace{-0.6 cm}
\end{figure}

The values mentioned above are used to assess the correctness of predicted pixels. As shown in  Table \ref{integrity}, our methods utilizing various backbones obtain state-of-the-art FNR results on 3 out of 5 datasets, showcasing superior integrity learning capabilities on most benchmarks. Example visualizations in Fig. \ref{intergrity_img} further highlight the improved integrity of our polyp segmentation.

\begin{table}[htbp]
  \vspace{-5 mm}
  \centering
  \caption{FNR of different methods in different datasets.}
    \resizebox{\linewidth}{!}{
        \begin{tabular}{cccccc}
        \toprule
        \multicolumn{1}{c|}{Methods} & CVC-300 $\downarrow$ & ClinicDB & $\downarrow$ ColonDB $\downarrow$ & ETIS $\downarrow$  & Kvasir-SEG $\downarrow$ \\
        % \midrule
        \midrule
        \multicolumn{1}{c|}{Unet} & 0.231 & 0.163 & 0.483 & 0.526 & 0.127 \\
        \multicolumn{1}{c|}{Unet++} & 0.269 & 0.206 & 0.513 & 0.612 & 0.177 \\
        \multicolumn{1}{c|}{SFA} & 0.108 & 0.196 & 0.295 & 0.366 & 0.182 \\
        \multicolumn{1}{c|}{PraNet} & 0.057 & 0.088 & 0.270  & 0.327 & \textbf{0.075} \\
        \multicolumn{1}{c|}{ACSNet} & 0.032 & 0.084 & 0.234 & 0.252 & 0.083 \\
        \multicolumn{1}{c|}{DCRNet} & 0.051 & 0.081 & 0.215 & 0.244 & 0.078 \\
        \multicolumn{1}{c|}{MSEG} & 0.066 & 0.076 & 0.250  & 0.264 & 0.087 \\
        \multicolumn{1}{c|}{SANet} & 0.034 & 0.056 & 0.202 & \textbf{0.107} & 0.081 \\
        \midrule
        \multicolumn{1}{c|}{IC-PolypSeg-EF0} & 0.049 & 0.065 & 0.237 & 0.230  & 0.088 \\
        \multicolumn{1}{c|}{IC-PolypSeg-EF1} & 0.047 & 0.060  & 0.218 & 0.224 & 0.091 \\
        \multicolumn{1}{c|}{IC-PolypSeg-EF2} & 0.046 & 0.062 & 0.195 & 0.176 & 0.085 \\
        \multicolumn{1}{c|}{IC-PolypSeg-EF3} & 0.049 & 0.061 & 0.213 & 0.195 & 0.092 \\
        \multicolumn{1}{c|}{IC-PolypSeg-EF4} & 0.067 & 0.065 & 0.208 & 0.176 & 0.102 \\
        \multicolumn{1}{c|}{IC-PolypSeg-EF5} & 0.063 & 0.065 & 0.203 & 0.167 & 0.104 \\
        \multicolumn{1}{c|}{IC-PolypSeg-EF6} & 0.050  & 0.049 & \textbf{0.170} & 0.138 & 0.080 \\
        \multicolumn{1}{c|}{IC-PolypSeg-EF7} & \textbf{0.032} & \textbf{0.042} & 0.178 & 0.145 & 0.078 \\
        \midrule
        \end{tabular}
        }
  \label{integrity}%
  \vspace{-6 mm}
\end{table}%

\vspace{-2mm}
\subsection{Quality Comparisons}
\begin{figure}
\centering
\includegraphics[width=0.9\linewidth]{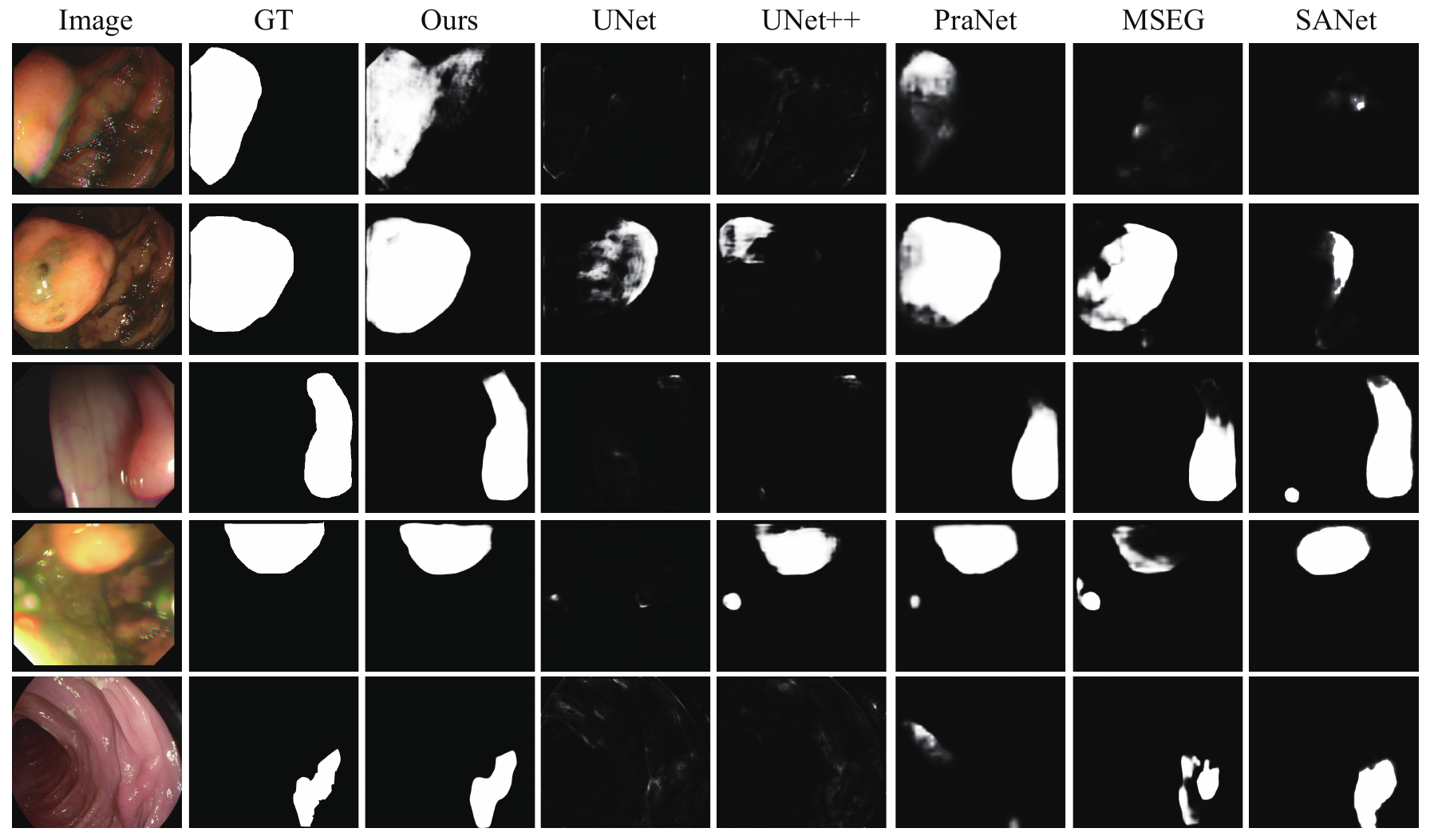}
\caption{Visual comparisons of our proposed method (IC-Polyp-EF7) with other five SOTA methods.}
\label{results}
\end{figure}

Fig. \ref{results} provides quality comparisons between our proposed method and other five SOTA methods. As we can observed, Our proposed method (IC-PolypSeg-EF7) generates more accurate polyp segmentation results for various cases. The results demonstrate our method segments polyp regions with less noise and greater integrity compared to other SOTA methods. 

\vspace{-3mm}

\subsection{Ablation Study}

To validate the effectiveness of the proposed PSFR, CPSFR and CFC modules, we conducted ablation studies as shown in Table \ref{ablation}. The baseline model (ID:1) contains the IC-PolypSeg-EF7 and the RFE module. The PSFR module (ID:2) leads to performance gains, demonstrating its contribution. Further integrating the CPSFR modules (ID:3) provides further boosts by capturing pixel-wise contextual information. Finally, combining all three modules (ID:4) achieves the overall best results, confirming that each proposed module contributes positively to the framework.

\vspace{-4mm}

\begin{table}[htbp]
  \centering
  \caption{Ablation analysis of gradually adding the proposed modules. The best performance are shown in bold. }
  \resizebox{\linewidth}{!}{
    \begin{tabular}{c|c|cc|cc|cc|cc|cc}
    \toprule
    \multirow{2}[2]{*}{ID} & \multirow{2}[2]{*}{Componet Settings} & \multicolumn{2}{c|}{EndoScene} & \multicolumn{2}{c|}{ClinicDB} & \multicolumn{2}{c|}{Colon-DB} & \multicolumn{2}{c|}{ETIS} & \multicolumn{2}{c}{Kvasir-SEG} \\
          &       & Dice $\uparrow$  & IoU $\uparrow$   & Dice $\uparrow$  & IoU $\uparrow$   & Dice $\uparrow$  & IoU $\uparrow$   & Dice $\uparrow$  & IoU $\uparrow$   & Dice $\uparrow$  & IoU $\uparrow$ \\
    \midrule
    1     & Baseline   & 0.886 & 0.815 & 0.915 & 0.858 & 0.746 & 0.659 & 0.702 & 0.617 & 0.884 & 0.832 \\
    2     & + PFR & 0.895 & 0.828 & 0.923 & 0.869 & 0.767 & 0.686 & 0.731 & 0.654 & 0.889 & 0.837 \\
    3     & + PFR + CPFR & 0.903 & 0.839 & 0.932 & 0.878 & 0.787 & 0.699 & 0.745 & 0.663 & 0.893 & 0.848 \\
    4     & + PFR + CPFR + CFC & \textbf{0.907} & \textbf{0.846} & \textbf{0.938} & \textbf{0.890}  & \textbf{0.799} & \textbf{0.728} & \textbf{0.758} & \textbf{0.670}  & \textbf{0.910}  & \textbf{0.859} \\
    \bottomrule
    \end{tabular}%
    }
  \label{ablation}%
\end{table}%

\vspace{-4mm}

\section{Conclusion}

In this work, we have introduced a Integrity Capturing Polyp Segmentation network (IC-PolypSeg), which accurately segments polyp regions in colonoscopy images while preserving prediction integrity. IC-PolypSeg integrates three novel modules, PSFR, CPSFR, and CFC, achieving a balance between accuracy, integrity, and efficiency. Comprehensive experiments on five public datasets demonstrate the efficacy of each proposed module, achieving new state-of-the-art results. Moreover, these modules exhibit strong potential for extension to other medical imaging domains like CT, MRI, and histology. We aim to explore these extensions in future work.

\section{Acknowledgements}
This work was supported by grants from the National Natural Science Foundation of China (No. 82171837 to Yun Liu), the National Key R\&D Program of China (No. 2021YFC2701001 to Yun Liu), Program of Shanghai Academic/Technology Research Leader Grant (No. 20XD1420400 to Yun Liu) and Shanghai Municipal Science and Technology Major Project (No. 2018SHZDZX01) and ZJLab.

% \small
\bibliographystyle{IEEEbib}
\bibliography{strings,ms}

\end{document}